# PARALLEL PERFORMANCE OF MPI SORTING ALGORITHMS ON DUAL–CORE PROCESSOR WINDOWS-BASED SYSTEMS

Alaa Ismail El-Nashar

Faculty of Science, Computer Science Department, Minia University, Egypt
Assistant professor, Department of Computer Science, College of Computers and Information Technology, Taif University, Saudi Arabia

```
a.ismail@tu.edu.sa
nashar_al@yahoo.com
```

## Abstract

*Message Passing Interface (MPI) is widely used to implement parallel programs. Although Windows-based architectures provide the facilities of parallel execution and multi-threading, little attention has been focused on using MPI on these platforms. In this paper we use the dual core Window-based platform to study the effect of parallel processes number and also the number of cores on the performance of three MPI parallel implementations for some sorting algorithms.*

## 1. INTRODUCTION

There are three main models for parallel programming multi-core architectures. These models are the message-passing paradigm (MPI), shared memory programming model, and Partitioned Global Address Space (PGAS) programming model [4].
Message Passing Interface (MPI) [20] is the most commonly used paradigm in writing parallel programs since it can be employed not only within a single processing node but also across several connected ones. MPI standard has been designed to enhance portability in parallel applications, as well as to bridge the gap between the performance offered by a parallel architecture and the actual performance delivered to the application [5]. Two critical areas determine the overall performance level of an MPI implementation. The first area is the low-level communication layer that the upper layers of an MPI implementation can use as foundations. The second area covers the communication progress and management [5].
MPI offers several functions such as point-to-point rendezvous-type send/receive operations, logical process topology, data exchange, gathering and reduction operations to combine partial results from parallel processes, and synchronization capabilities manifested in barrier and event operations.
The shared memory programming model allows a simpler programming of parallel applications, as the control of the data location is not required. OpenMP [6] is the most widely used solution for shared memory programming, as it allows an easy development of parallel applications through compiler directives. Moreover, it is becoming more important as the number of cores per system increases. However, as this model is limited to shared memory architectures, the performance is bound to the computational power of a single system. To avoid this limitation, hybrid systems, with both shared/distributed memory, such as multi-core clusters, can be programmed using MPI combined OpenMP. However, this hybrid model can make the parallelization more difficult and the performance gains could not compensate for the effort [4].

Partitioned Global Address Space (PGAS) programming model combines the main features of the message passing and the shared memory programming models. In PGAS languages, each thread has its own private memory space, as well as an associated shared memory region of the global address space that can be accessed by other threads, although at a higher cost than a local access. Thus, PGAS languages allow shared memory-like programming on distributed memory systems. Moreover, as in message-passing, PGAS languages allow the exploitation of data locality as the shared memory is partitioned among the threads in regions, each one with affinity to the corresponding thread.

Several implementations such as Parallel Virtual Machine (PVM) [17] and MPICH2 [19] are now available and can be used in writing MPI programs. Parallel Virtual Machine (PVM) [17], is a software package that permits a heterogeneous collection of UNIX or Windows computers hooked together by a network to be used as a single large parallel computer.
MPICH2 is a high performance and widely portable implementation of MPI standard. It efficiently supports different computation and communication platforms. It also supports using of C/C++ and FORTRAN programming languages. In contrast to MPICH2 for Windows, the implementation for UNIX and LINUX offers built-in network topology support. This makes an easy use of MPICH2 on such platforms and hence little attention has been focused on using the implementation on Microsoft Windows although it provides the facilities of parallel execution and multi-threading.

MPICH2 for windows can be installed either on a single machine having single / multi-core processors or an interconnected set of machines. In both cases, performance of MPI programs is affected with various parameters such the number of cores (machines), number of running processes and the programming paradigm which is used.

The construction of MPICH emphasizes its Unix origins. It uses only one thread per MPI process and works with one type of communication medium at a time. These design characteristics also helped the portability of the library, since not all platforms have threads. Although, some mechanisms were implemented to use more than one device, in fact they were very poor and hard to use. Since only one thread was available, the library had to make polling on the devices to get new messages. In addition, the message passing progress could only be possible when the user thread made a call to a MPI function. This architecture's type is well fitted for super computers or dedicated clusters where there is only one process per CPU. However, the typical NT cluster is shared among several users, which may be executing interactive tasks.

The goals of MPICH2 are to provide an MPI implementation that efficiently supports different computation platforms including commodity clusters such as desktop systems, shared-memory systems, and also multi-core architectures. It also supports and different communication platforms including high-speed networks such as 10 Gigabit Ethernet, InfiniBand, and Myrinet.
In this paper we focus on using MPICH2 for Windows on a dual-core processor machine to study how the number of cores and also the number of processes affect the performance.
Three parallel sorting algorithms namely Bubble sort, Merge sort and Quick sort are designed and implemented using MPI. The effect of the number of cores and also the number of processes on the algorithms performance is studied.
The paper is organized as follows: section 2 gives a brief idea about the related work. In section 3, we discuss sequential sorting algorithms and parallelization methods. Section 4 presents the experiments carried out and the gained results.

## 2. RELATED WORK

Chip Multiprocessors (CMP) [8] is a multithreaded architecture, which integrates more than one processor on a single chip. In this architecture, each processor has its own L1 cache. The L2 cache and the bus interface are shared among processors. Intel Core 2 Duo [7] is an example of such architecture; it has two processors on a single chip, each of them has an L1 cache, and both of them are sharing the L2 cache.

These architectures not only provide a facility for implementing and running the parallelized applications without a need for building interconnected machines but also enhance the data management operations among parallel processes due to the reliable utilization of hardware resources.

Multi-core architectures are designed to provide a high performance feature on a single chip since that they do require neither a complex system nor increased power requirements [10].

On the other hand, many parameters such as latency, bandwidth, caches and even the system software [14] affect the performance of such systems. These challenges should to be studied to gain the objective of these architectures.

Several Studies [4], [15], [16] , [18] have been addressed the performance of MPI applications on several hardware platforms, but little attention has been focused on using multi-core architectures supported by Microsoft Windows as an operating system and MPICH2 as an MPI implementation.

In this paper we use the dual core Window-based platform to study the effect of parallel processes number and also the number of cores on the performance of three MPI parallel implementations for some sorting algorithms.

## 3. SORTING ALGORITHMS

The main function of sorting algorithms is to place data elements of a list in a certain order. Several algorithms are introduced to solve this problem.

### 3.1 Sequential sorting algorithms

Sequential sorting algorithms are classified into two categories. The first category, "distribution sort", is based on distributing the unsorted data items to multiple intermediate structures which are then collected and stored into a single sorted list.

The second one, "comparison sort", is based on comparing the data items to find the correct relative order [9].

In this paper we focus on comparison based sorting algorithms. These algorithms use various approaches in sorting such as exchange, partition, and merge.

The exchange approach repeats exchanging adjacent data items to produce the sorted list as in case of bubble sort [11].

The partitioning approach is a "divide and conquer" strategy based on dividing the unsorted list into two sub-lists according to a pivot element selected from the list. The two sub-lists are sorted and then combined giving the sorted list as in case of quick sort [13].

Merge approach is also a divide and conquer strategy that does not depend on a pivot element in portioning process. The approach repeatedly divides the original list into sub-lists until the sub-lists have only one data item. Then these elements are merged together given the sorted list as in case of merge sort [2], [3].

### 3.2 Parallelizing sorting algorithms

Parallelizing sorting algorithms needs a careful design to achieve well efficient results because of the high level date dependency evolved within these algorithms that exhibits parallelism.

Sequential versions of bubble sort, quick sort and merge sort are parallelized using C ++ binding of MPI under MPICH2 for Windows. The "scatter/ merge" paradigm is used in parallelization.

The used paradigm has three fundamentals phases, scatter phase, sort phase and merge phase. The first phase is responsible for distributing the original unsorted data list among the MPI process in such a way each of them accepts a part of the original data to be manipulated with these parallel processes.

In sort phase, each process sorts its local unsorted data list using one of the selected sorting algorithms. All local sorted data are sent from these "slave" processes to only one process which serves as a master process to generate the sorted list in the merge phase. An MPI skeleton of this paradigm is shown in figure 1.

```
1. Initialize MPI environment.
2. Determine the number of MPI processes (p) and their id's.
/*                Scatter Phase                    */
3. If  id=master then
4.         get unsorted list data  items of size n
5.         compute partition size, s = n/p
6.         broadcast s and n to all processes
7. endif
8. scatter sub-lists from master process to all running processes
/*                Sorting Phase                    */
9. call Selected_Sorting_Algorithm ( sub-list, s)
/*                Merge Phase                      */
10.  while step< p do
11.      if   id is even then
12.          Send even-sub-list to process id + 1
13.          Receive odd-sub-list from processor id + 1
14.          Merge even-sub-list and odd-sub-list into sorted-list
15.          Replace even-sub-list by the first half of sorted-list
16.      else if id > 0 then
17.          Receive even-sub-list from process id - 1
18.          Send odd-sub-list to process id - 1
19.          Merge even-sub-list and odd-sub-list into sorted-list
20.          Replace odd-sub-list by the second half of sorted-list
21.      end if
22. End while
23. Finalize MPI environment
24. End
```

Figure 1. MPI scatter/ merge paradigm

### 3.2.1 Parallelizing bubble sort algorithm

Sequential version of bubble sort [11] is a simple sorting algorithm. It repeats exchanging adjacent data items to produce the sorted list.
We implemented a parallel MPI version of bubble sort using scatter/ merge paradigm as shown in figure 1.

In sorting phase all parallel processes sort their local sub-list with sizes s using sequential bubble sort algorithm that uses elements exchanging function as described in figure 2.

```
/*                    Bubble Sort                          */
1. Bubble ( sub-list, s)
2.    { define counters  i and j
3.        for (i = s - 2; i >= 0; i--)
4.          for (j = 0; j <= i; j++)
5.            if (sub-list [j] > sub-list[j + 1])
6.              swap (sublist, j, j + 1);
7.    }

/*                  Exchange_Function                      */

1. Swap (sub-list [], int i, int j)
2.    {
3.      int temp;
4.      temp = sub-list[i];
5.      sub-list[i] = sub-list[j];
6.      sub-list[j] = temp;
7.    }
```

Figure 2. Bubble sort and Exchange

### 3.2.2 Parallelizing merge sort algorithm

Like quick sort, the sequential merge sort algorithm uses a "divide and conquer" strategy to sort an unsorted data items list. The difference between the two algorithms is that merge sort does not depend on selecting a pivot element. The original list is repeatedly divided into two equal size sub-lists until each sub-list contains a single data element. These elements are then merged together as pairs to generate sorted sub-lists having only two data elements per each sub-list. Figure 3 shows an illustration of the parallel merge sort algorithm.

```
/*                    Merge Sort                           */
 1. merge_sort(sub-list[], start, last)
 2. {  Allocate spaces for "sublist1" and "sub-list2" of size (last-start)/2 each;
 3.    mid = (first+last)/2;
 4.    lcount = mid - first + 1;
 5.    ucount = last - mid;
 6.    if (last == first) { return;
 7.      } else {
 8.          sub-list1=merge_sort(sub-list[], first, mid);
 9.          Sublist2=merge_sort(sub-list[], mid+1, last);
10.          merge(sublist1, lcount, sublist2, ucount);
11.      }
12. }
```

Figure 3. Merge sort implementation outline

The generated sub-lists are then merged until the sorted list having the original number of data elements is generated.
An MPI parallel version of this algorithm is implemented by partitioning the original list into parts having the same size. One of parallel processes is designated as a master. This process distributes the data parts among other workers parallel processes that use the sequential version of merge sort to sort their own data. The sorted sub-lists are sent to the master. Finally, the master merges all the sorted sub-lists into one sorted list.

### 3.2.3 Parallelizing quick sort algorithm

Sequential quick sort is a "divide and conquer" algorithm that sorts an unsorted data items list by recursively partitioning it into smaller sub-lists. The algorithm selects a data item," pivot element", from the unsorted list. Then the list is re-ordered in such a way that all items with values less than the pivot are placed to the left of the pivot, and all items with values greater than the pivot are placed to its right. The same operation is executed recursively to sort the elements of the two "right" and "left" sub-lists

Sorting these smaller sub-lists can be carried out simultaneously since there is no data dependency among them. This gives a high opportunity of parallelism.

We implemented an MPI parallel version of quick sort that is also based on "scatter/merge" paradigm as shown in case of bubble sort. The main difference between the two implementations is the sorting algorithm used in sorting phase.

In parallel quick sort MPI implementation, the master process broadcasts the original list size, to all processes. Also the unsorted list is scattered among the processes which all apply quick sort algorithm on their own lists, the outline of sorting algorithm used in sorting phase is shown in figure 4.

```
/*                        Quick Sort                          */

// (quick) sort slice of array v; slice starts at s and is of length n
 1. quicksort(sub-list v, start, length)
 2. {
 3.   int pivot, pstart, i;
 4.   if (length <= 1)
 5.     return;
    // pick pivot and swap with first element
 6.   pivot = sub-list[start + length/2];
 7.   swap(sub-list[], start, s + length/2);
    // partition slice starting at s+1
 8.   pstart = start;
 9.   for (i = start+1; i < start+length; i++)
10.   if (sub-list[i] < pivot)
11.     { pstart++;
12.       swap(sub-list[], i, pstart);
13.     }
    // swap pivot into place
14.   swap(sub-list[], start, pstart);
    // recurse into partition
15.   quicksort(sub-list[], start, pstart-start);
16.   quicksort(sub-list[], pstart+1, start+length-pstart-1);
17. }
```

Figure 4. Quick sort implementation outline

### 3.3 Parallelization challenges

All of the sequential algorithms discussed above are challenged by various factors such as computation complexity and memory usage. Parallelizing these algorithms adds several extra challenges such as application speed up and how can it be affected by the number of cores and/ or the number of the running processes.

In sequential bubble sort described in figure 2, the computation complexity is $O(n^2)$, $n$ is the unsorted list size, in both best and average case [11].

The parallel version complexity of bubble sort based on "scatter/ merge" paradigm is estimated as $O(n^2/P^2)$, so the time will be reduced by a factor of $P^2$, $P$ is the number of processors. This is due to the partitioning of the total size n of the original list among the running processes $P$. This implies a theoretical super linear speed up. In case of using only two cores we expect that the computation complexity of parallel bubble sort will be $O(n^2/mP)$, $m$ is the number of processes and P is the number of physical cores. This implementation uses a set of parallel processes linearly; each process communicates only with its two neighbors; this makes a negligible communication overhead. Also it does not require excessive memory locations since each process manipulates always $n/p$ data items, and all buffers can be exactly allocated at the beginning of the execution so we expect that its memory usage complexity will be $O(n) + O(1)$, $n$ is the data size. This means that it consumes a fixed memory usage beside the original size of the list.

Sequential merge sort algorithm behaves as $O(n \log n)$ computational complexity in all of its cases, worst, average and best [12]. We estimated the parallelized version complexity in case of using a dual core processor as $\frac{2}{P}\log(n/m) + total\ overhead$. The total overhead is the sum of inter-process communications overhead and the MPI processes initialization overhead.

As shown in figure 3, the algorithm uses duplicate list size memory locations; the extra locations are needed in merging the sorted sub-lists.

In case of sequential quick sort described in figure 4, the efficiency of the algorithm is influenced the pivot element selection method; we get worst case $O(n^2)$ when the selected pivot is the left most data item. If the pivot is carefully selected, the algorithm behaves in its best case as $O(n \log n)$ complexity [12]. In case of parallelized version, we estimated the complexity for a dual core processor as $\frac{2n}{mP}\log(n/m) + total\ overhead$. Parallel quick sort uses a hypercube topology of processes; each process exchanges data items with its $\log P$ neighbors. We predicted that the implementation will use $\log n$ memory size in addition to the space used to store the original list.

## 4. EXPERIMENTAL PERFORMANCE

Two experiments are carried out and applied to the entire implemented parallel version of the concerned sorting algorithms.

"Experiment 1" is designed to address the affect of parallel processes number, and also the number of used cores on the performance. "Experiment 2" is designed to detect whether the theoretical memory usage complexity is compatible with experimental results. The outline of the experiments is summarized below.

**Experiment 1**

1. Set the number of system cores to 1 and reboot the system.

International Journal of Distributed and Parallel Systems (IJDPS) Vol.2, No.3, May 2011

2. Execute the parallel MPI application on the same single core repeatedly using arbitrary number of MPI processes, 1, 2, 3... , n for the same data with the same size.
3. Record execution time.
4. Set the number of system cores to 2 and reboot the system.
5. Repeat steps 2-4 with the same data and size.

**Experiment 2**

1. If number of cores = 1 set it to 2 and reboot the system.
2. Execute the parallel MPI application using only two MPI processes with an arbitrary data of an arbitrary size.
3. Rerecord memory usage.
4. Increase data size.
5. Repeat steps 2-4.

We used an experimental system consists of Pentium[R] Dual-Core CPU E5500@ 2.80 GHZ, 3.21 GB of RAM running on Microsoft Windows XP Professional Service Pack 2. The experiments codes were written in C++ using MPICH2 version 1.0.6p1, as a message passing implementation.

### 4.1 Results of Experiment 1

We applied "Experiment 1" to the three parallel implementations with a fixed data size $2\times10^5$ for bubble sort and $6\times10^6$ for merge sort and quick sort respectively with 1,2, ..64 parallel process using both a single and dual cores as shown in table 1.

Table 1. Results of Experiment 1

| Number of cores | Number of processes | Execution time in seconds | | |
|---|---|---|---|---|
| | | Bubble sort, $2\times10^5$ date items | Merge sort, $6\times10^6$ date items | Quick sort, $6\times10^6$ Date items |
| 1 | 1 | 457.375 | 5.718 | 4.437 |
| | 2 | 229.250 | 5.765 | 4.500 |
| | 4 | 115.859 | 5.906 | 4.562 |
| | 8 | 57.812 | 6.296 | 4.859 |
| | 16 | 28.953 | 6.953 | 5.421 |
| | 32 | 15.140 | 8.421 | 6.890 |
| | 64 | 9.0460 | 11.687 | 10.14 |
| 2 | 1 | 460.796 | 5.609 | 4.203 |
| | 2 | 117.546 | 4.031 | 3.203 |
| | 4 | 57.937 | 4.234 | 3.328 |
| | 8 | 29.109 | 4.328 | 3.453 |
| | 16 | 14.646 | 4.640 | 3.796 |
| | 32 | 7.625 | 4.968 | 4.265 |
| | 64 | 4.687 | 7.109 | 6.265 |

We profiled the execution of the tested implementations using jumpshot [1] to address the inter-processes communication. Also the total overhead and computation costs are measured.
As the theoretical expectation, the execution time of bubble sort is reduced as the number of parallel processes increases in case of using either single or dual cores as shown in figure 5. On other hand merge sort and quick sort do not exhibit a speed up behavior as processes number increases as shown in figure 6 and figure 7.

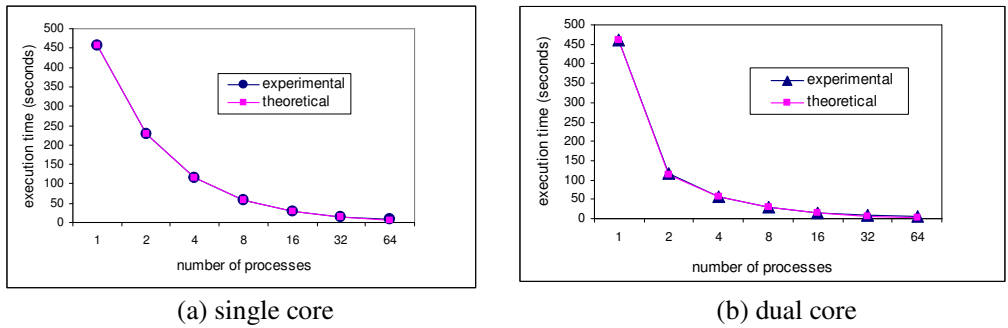

(a) single core    (b) dual core

Figure 5. Experimental and theoretical execution time of parallel bubble sort

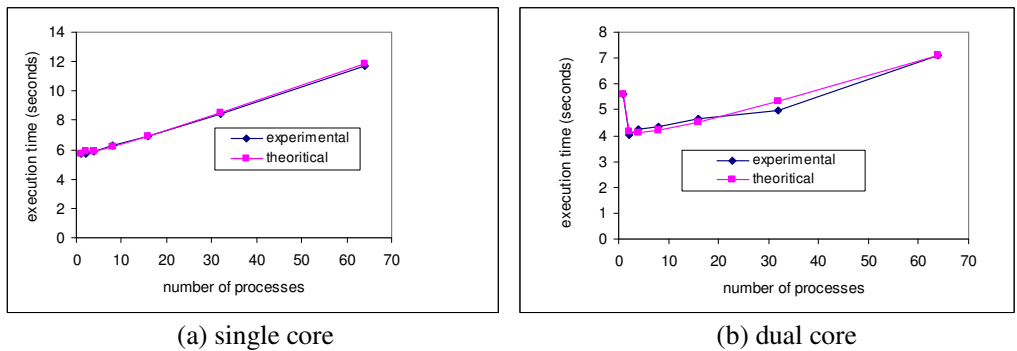

(a) single core    (b) dual core

Figure 6. Experimental and theoretical execution time of parallel merge sort

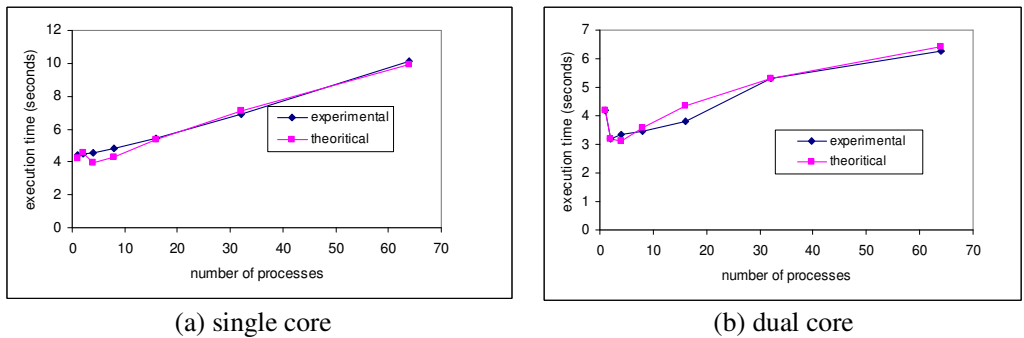

(a) single core    (b) dual core

Figure 7. Experimental and theoretical execution time of parallel quick sort

To interpret these result, we profiled the execution of the tested implementations using jumpshot to address the inter-processes communication. Also the total overhead and computation costs are measured. Figure 8 shows how the running parallel processes communicate with each others. In bubble sort (figure 8.a) there is a low communication overhead compared with that of computations in contrast to figures 8.b and 8.c that show a higher communication overhead. The excessive inter-process communications overhead noticed in both merge sort and quick sort increases the total execution time.

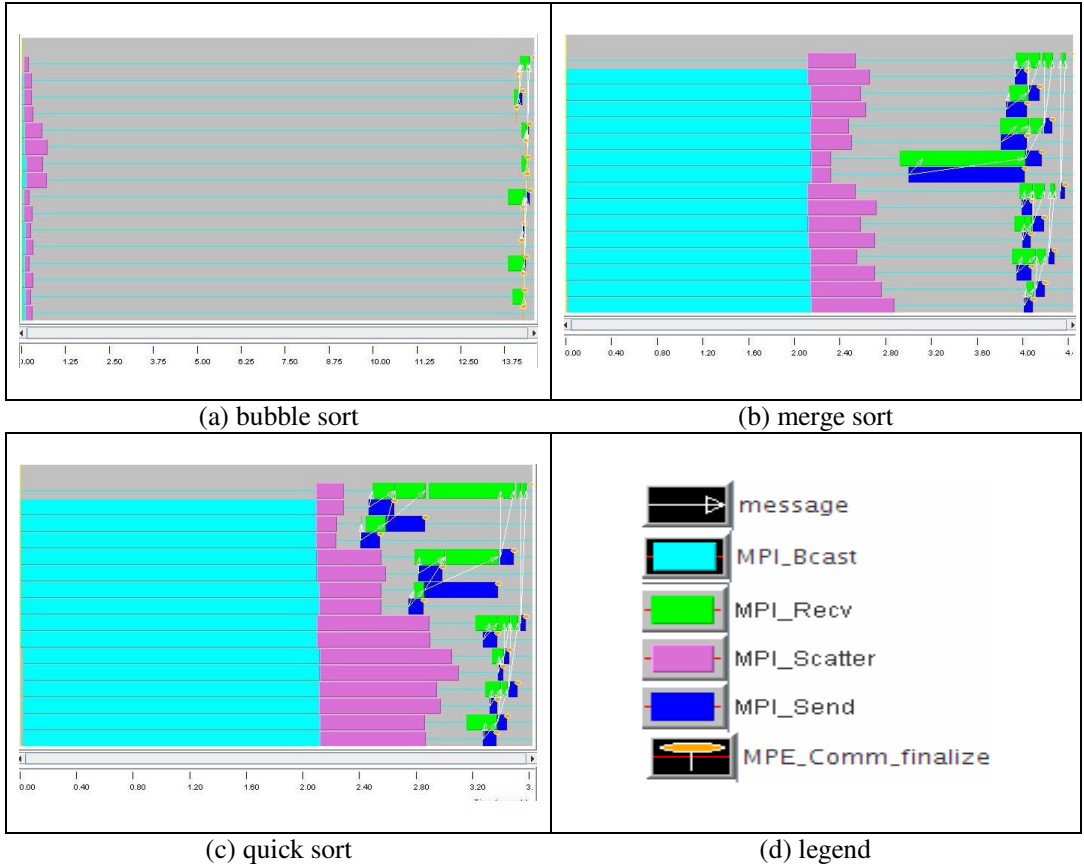

Figure 8. Jumpshot time line for experiment 1

We also measured the total overhead of parallel processes compared with the computation cost for the three implementations regarding the number of processes and also the number of cores used as shown in figure 9 , figure 10 and figure 11.

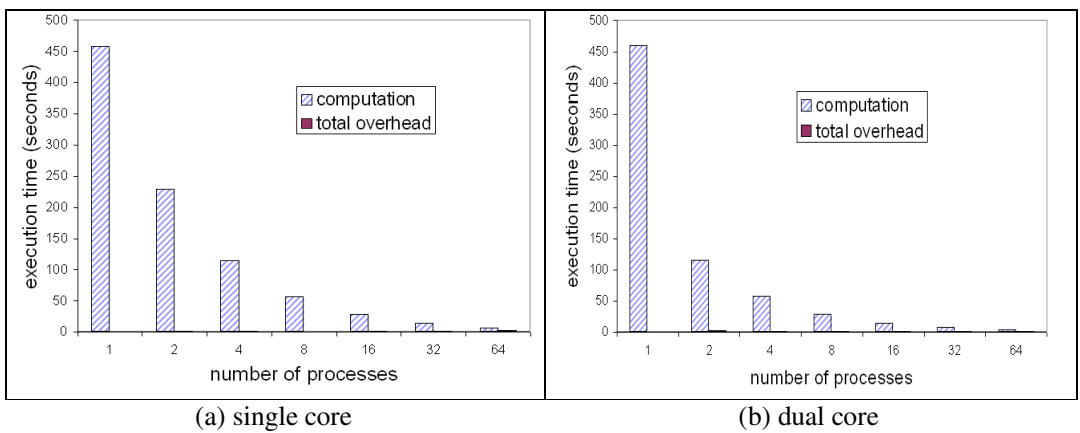

Figure 9. Bubble sort overhead/ computation ratio

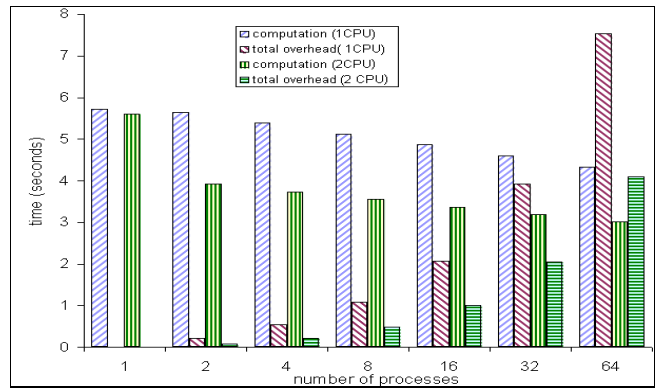

Figure 10. Merge sort overhead/ computation ratio

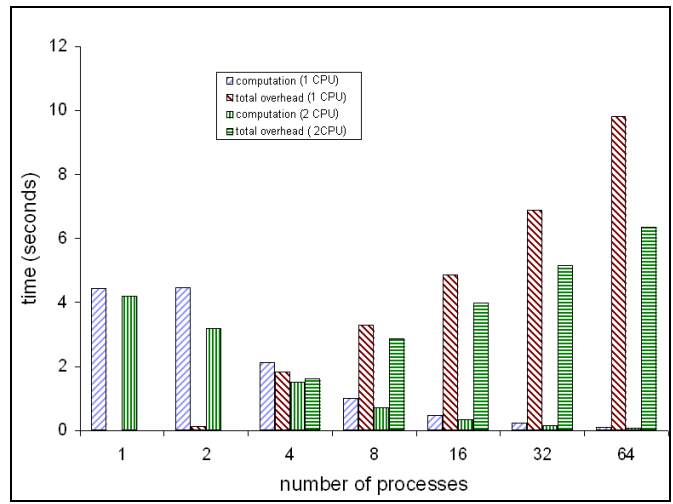

Figure 11. Quick sort overhead/ computation ratio

## 4.2 Results of Experiment 2

We applied "Experiment 2" to the three parallel implementations with only two parallel processes running on two cores and different data sizes. The memory usage is recorded as shown in Table 2.

Table 2. Results of Experiment 2

| Memory usage in Mega Bytes. | | | | |
|---|---|---|---|---|
| Data size x $10^4$ | Bubble sort | Data size x $10^6$ | Merge sort | Quick Sort |
| 5 | 4.984 | 2.5 | 204.560 | 38.976 |
| 6 | 5.104 | 3 | 248.412 | 45.824 |
| 7 | 5.224 | 3.5 | 289.692 | 52.684 |
| 8 | 5.344 | 4 | 331.076 | 59.548 |
| 9 | 5.456 | 4.5 | 370.740 | 66.400 |
| 10 | 5.576 | 5 | 414.260 | 73.256 |

For bubble sort, the data size is increased from $5\times 10^4$ to $1\times 10^5$ data items stepping $10^4$ data items in each run. For the other two implementations, the data size is increased from $2.5\times 10^6$ to $1\times 10^5$ data items stepping $5\times 10^5$ data items in each run. Figure 12 shows these experimental results compared with the theoretical ones.

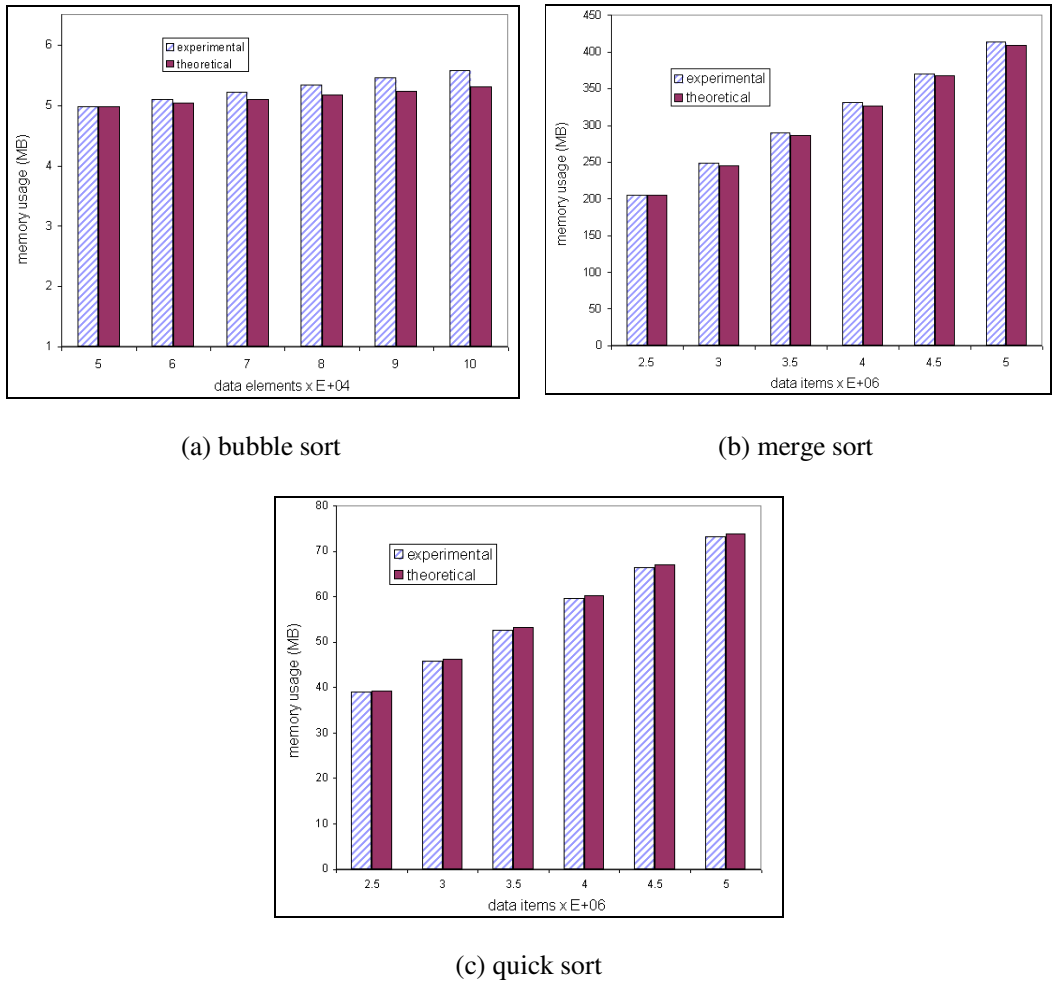

(a) bubble sort  (b) merge sort

(c) quick sort

Figure 12. Experimental and theoretical memory usage

## 5. CONCLUSION

In this paper we focused on using MPICH2 as a message passing interface implementation on Windows platforms. The effect of parallel processes number and also the number of cores on the performance of parallel bubble sort, parallel merge sort and parallel quick sort algorithms has been theoretically and experimentally studied.

We found that the computation/ communication ratio greatly affects the execution time of the three studied parallel algorithms.

Although bubble sort algorithm is the slowest one compared with merge sort and quick sort, its execution time decreased rapidly as the number of processes increased even the number of processes is greater than the number of physical cores. This is because it does not require heavy communication; the great part of the execution time is consumed in computations. So as the number of processes increases the amount of work done by each process will be decrease regardless the effect of the number of physical cores used. In contrast to this situation the execution times of merge sort and quick sort were very sensitive to both the number of running processes and the number of cores used. The execution times for both of them increase as the number of processes exceeds the number of cores. The total overhead generated from processes initialization and inter-process communication negatively affects the execution time.

Although quick sort is the fastest one among the three algorithms, it suffers form a high communication overhead cost and load imbalance compared with merge sort.

We also compared the memory usage for the three algorithms theoretically and experimentally. Our estimation of memory usage was very close to the experimental results.

International Journal of Distributed and Parallel Systems (IJDPS) Vol.2, No.3, May 2011

**Author**

**Alaa I. Elnashar** was born in Minia, Egypt, on November 5, 1967. He received his B.Sc. and M.Sc. from Faculty of Science, Department of Mathematics (Math. & Comp. Science), and Ph.D. from Faculty of Science, Department of Computer Science, Minia University, Egypt, in 1988, 1994 and 2005. He is a staff member in Faculty of Science, Computer Science Dept., Minia University, Egypt.

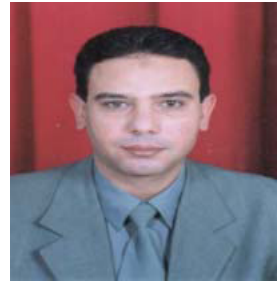

Dr. Elnashar was a postdoctoral fellow at Kanazawa University, Japan. His research interests are in the area of Software Engineering, Software Testing, and parallel programming.

Now, Dr. Elnashar is an Assistant professor, Department of Computer Science, College of Computers and Information Technology, Taif University, Saudi Arabia.